\documentclass[twocolumn,showpacs,preprintnumbers,prl]{revtex4}
\usepackage{graphicx}
\usepackage{dcolumn}
\usepackage{amssymb}
\usepackage{bm}

\begin{document}

\title{Acoustic Wave Absorption as a Probe of Dynamical Geometrical Response of Fractional Quantum Hall Liquids}
\author{Kun Yang}

\affiliation{NHMFL
and Department of Physics, Florida State University, Tallahassee,
Florida 32306, USA}

\begin{abstract}
We show that acoustic crystalline wave gives rise to an effect similar to that of a gravitational wave to an
electron gas. Applying this idea to a two-dimensional electron gas in the fractional quantum Hall regime, this
allows for experimental study of its intra-Landau level dynamical response in the long-wave length limit. To
study such response we generalize
Haldane's geometrical description of fractional quantum Hall states to situations where the external metric is
time-dependent. We show that such time-dependent metric (generated by acoustic wave) couples to collective
modes of the system, including a quadrapolar mode at long wave length, and magneto-roton at finite wave length.
Energies of these modes can be revealed in spectroscopic measurements.
controlled by strain-induced Fermi velocity anisotropy.
We argue that such geometrical probe provides a potentially highly useful alternative probe of quantum Hall
liquids, in addition to the usual electromagnetic response.
\end{abstract}

\date{\today}

\maketitle

Fractional quantum Hall (FQH) liquid is the prototype topological state of matter. Haldane\cite{Metric} pointed
out recently that description of FQH liquids in terms of topological quantum field theories, while capturing the
universal and topological aspect of the physics, is incomplete in the sense that an internal geometrical degree
of freedom responsible for the intra-Landau level dynamics of the system is not included. This geometrical degree
of freedom, or internal metric, couples to anisotropy in the interaction between
electrons\cite{qiu,haowang,Apalkov} or the electron band structure\cite{boyang}, and its expectation value is
determined by energetics of the system. In a recent work\cite{yang} we showed that this internal metric parameter
manifests itself as anisotropy of composite fermion Fermi surface, which is measurable. Our quantitative result
compares favorably with recent experiments, in which electron mass anisotropy is induced and controlled by an
in-plane magnetic field\cite{gokmen,kamburov,kamburov2}. This demonstrates the observability of this internal
geometry.
It has also been argued\cite{Metric,boyang12,xiluo,golkar} that this internal metric may be viewed as a {\em
dynamical} degree of freedom, whose long-wave length dynamics corresponds to the collective excitations of the
system that can be viewed as ``gravitons"\cite{gravitonnote}.
In a parallel stream of works, much effort has been devoted to studying FQH liquids in a curved background
space\cite{son13,abanov,gromov,cho,can,gromov15,wu,bradlyn,can14,gromov152}, following earlier seminal work by
Wen and Zee\cite{WenZee}.

In the existing theoretical studies\cite{qiu,haowang,Apalkov,boyang,yang,murthy,Ghazaryan}, the background
geometry (or metric) provided by electron-electron interaction and/or band structure is static. The main purpose
of the present work is to generalize this to the case where the background metric is time-dependent, and show
that dynamics of the metric couples to the intra-Landau level collective modes of the FQH liquid. In particular,
we show that such time-dependent metric can be generated by acoustic waves, that play a role very similar to
gravitational wave in this context.
Such gravitational wave naturally couples to graviton and other collective modes of the system. This allows for
spectroscopic measurements of collective mode energies, in particular graviton energy, using acoustic wave
absorption.

Existing work on this subject\cite{qiu,haowang,Apalkov,boyang,yang,murthy,Ghazaryan} has thus far focused on
non-relativistic electrons. On the other hand graphene
has emerged as a new arena to study quantum Hall physics\cite{barlasreview}. A second purpose of the present work
is to show that much of the considerations in the present and earlier works carry over to Dirac electrons and
thus graphene straightforwardly, once we identify the anisotropy of Fermi velocity with the external metric.
We thus start our discussion below with a description of how the Fermi velocity anisotropy of Dirac electrons
translate into a background metric for FQH states they form.

{\em  External Metric of Dirac and Schrodinger Electrons} --
Consider the Hamiltonian
\begin{eqnarray}
H=T+V,
\end{eqnarray}
with the kinetic energy for massless Dirac electrons in a magnetic field taking the form
\begin{eqnarray}
T=\sum_j v^{\mu\nu}\sigma_\mu\Pi^j_\nu,
\label{eq:kinetic energy}
\end{eqnarray}
where $j$ is electron index, $\sigma_{\mu=1,2}$ are the Pauli matrices, and $v^{\mu\nu}$ is the (real) Fermi
velocity matrix. Repeated Greek indices are summed over.
\begin{eqnarray}
{\bf \Pi}={\bf p}+{e\over c}{\bf A}({\bf r})
\end{eqnarray}
is the mechanical momentum, $\nabla\times {\bf A}({\bf r})=-B\hat{z}$ thus the electrons move in a uniform
perpendicular magnetic field. The two components of ${\bf \Pi}$ satisfy the commutation relation
\begin{equation}
[\Pi_x, \Pi_y]=-\frac{i\hbar e}{ c}(\partial_x A_y - \partial_y A_x)= \frac{i\hbar eB}{c} = \frac{i\hbar^2}{
\ell^2},
\end{equation}
where $\ell=\sqrt{\hbar c/(eB)}$ is the magnetic length.

The easiest way to obtain the Landau level energies and corresponding wave functions is to square the kinetic
energy of a single electron:
\begin{eqnarray}
\left[v^{\mu\nu}\sigma_\mu\Pi_\nu\right]^2 = (v v^T)^{\alpha\beta}\Pi_\alpha \Pi_\beta - \frac{\hbar^2}{
\ell^2}(v^{11}v^{22}-v^{12}v^{21})\sigma_z,
\end{eqnarray}
from which it is clear that the zero energy Landau level (0LL, which will be the focus of the rest of this paper)
wave functions only have weight in one of the two components, and the symmetric matrix $vv^T$ plays a role
identical to the inverse effective mass matrix for quadratic bands:
\begin{eqnarray}
T=\frac{1}{2}(m^{-1})^{\mu\nu}\Pi_{\mu}\Pi_{\nu} = \frac{g^{\mu\nu}\Pi_{\mu}\Pi_{\nu}}{2m_0},
\label{eq:metric}
\end{eqnarray}
where $m^{-1}$ is the inverse effective mass tensor, $1/m_0$ is the geometric mean of the eigenvalues of
$m^{-1}$, and the (space-only) metric tensor $g$ is defined by the second equality above, which is symmetric and
unimodular.

For massless Dirac electrons, we may therefore diagonalize this matrix $v v^T$ to obtain its eigenvalues $av_F^2$
and $v_F^2/a$, with $\sqrt{a}v_F$ and $v_F^2/\sqrt{a}$ the Fermi velocities along the two principle directions,
defined to be the $x$ and $y$ directions hereafter, and $v_F$ is their geometric mean. $|a-1|$ is a measure of
the anisotropy. It is known\cite{Juan} that strain and ripple modifies the $v$ matrix in graphene, and thus $v^F$
and in particular the anisotropy $a$, which plays a role very similar to the effective mass anisotropy parameter
in a quadratic band (here the notation is the same as that of Ref. \onlinecite{yang}). In ordinary semiconductors
we expect strain of lattice also induces or modifies effective mass anisotropy. We thus discuss the massless
Dirac and (massive) Schrodinger electrons on equal footing in the remainder of the paper. In the notation of Eq.
(\ref{eq:metric}), a strain induces a change of metric and thus geometry (of spaces), and a time-dependent strain
plays a role similar to a gravitational wave, that can excite the ``gravitons" of the FQH systems as we will see
below.

{\em Geometrical Coupling of Intra-Landau Level Dynamics and External Metric} --
The intra-Landau level degrees of freedom are described by guiding center coordinates
\begin{eqnarray}
{\bf R}={\bf r}-(\ell^2/\hbar)\hat{z}\times{\bf \Pi},
\end{eqnarray}
which commute with ${\bf \Pi}$.
The interaction term
\begin{eqnarray}
V=\sum_{i < j} V({\bf r}_i-{\bf r}_j)=\frac{1}{2}\sum_{\bf q}V_{\bf q}\rho_{\bf q}\rho_{-{\bf q}},
\end{eqnarray}
where $V_{\bf q}$ is the Fourier transform of electron-electron interaction potential $V(r)$ (assumed to be
isotropic) and
\begin{eqnarray}
\rho_{\bf q}=\sum_{i}e^{i{\bf q}\cdot{\bf r}_i}
\end{eqnarray}
is the density operator.
In the large $B$ limit, Landau level spacing overwhelms $V$, and the electron motion is confined to a given
Landau level. In this case it is appropriate to project $V$ onto 0LL that results in a reduced Hamiltonian
involving the ${\bf R}$'s only\cite{yang}:
\begin{eqnarray}
\tilde{V}=\frac{1}{2}\sum_{\bf q}V_{\bf q}e^{-\frac{1}{2}(aq_x^2+q_y^2/a)\ell^2}\overline{\rho}_{\bf
q}\overline{\rho}_{-{\bf q}},
\label{eq:Vprojected}
\end{eqnarray}
where
\begin{eqnarray}
\overline{\rho}_{\bf q}=\sum_{i}e^{i{\bf q}\cdot{\bf R}_i}
\end{eqnarray}
is the guiding center density operator, and we choose $\hat{x}$ and $\hat{y}$ directions to be the diagonal
directions of $m^{-1}$ or $vv^T$, with anisotropy possibly induced by lattice distortion.
We note the {\em only} place that the background geometric parameter $a$ enters $\tilde{V}$ is in the Gaussian
form factor of 0LL. Once confined to the 0LL the difference between Dirac and Schrodinger electrons disappears,
and our discussions below apply to both.

\begin{figure}
\includegraphics[width=0.45\textwidth]{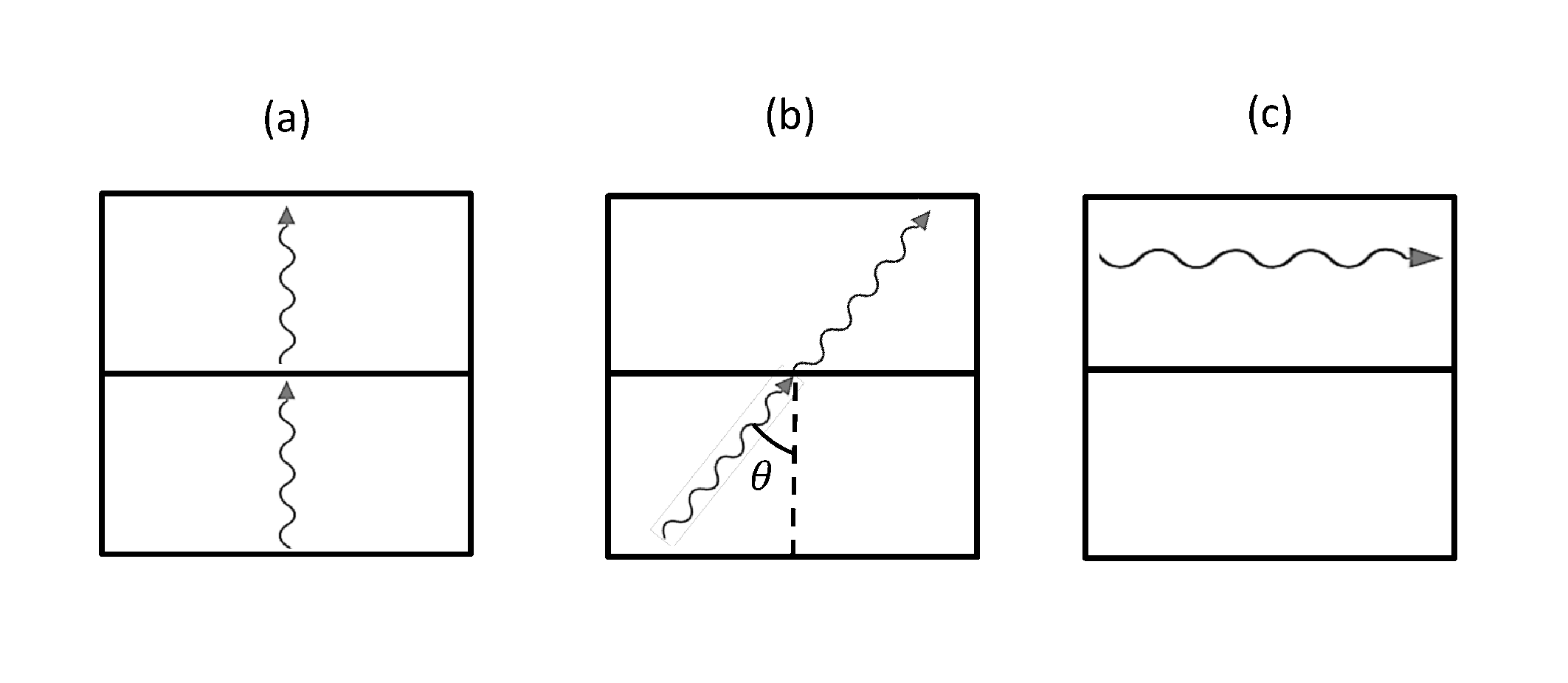}
\caption{Illustration of experimental setup. Square boxes represent 3D crystals, and thick solid lines represent
2D electron gas. Wavy lines represent acoustic waves whose effects are very similar to gravitational waves. (a)
Bulk acoustic wave propagating perpendicular to the 2D layer, inducing a uniform strain seen by 2D electron. (b)
Bulk acoustic wave propagating at an angle $\theta$ to the normal direction  of 2D layer, inducing a non-uniform
strain seen by 2D electron. (c) Surface acoustic wave propagating parallel to the 2D layer.}
\label{fig:Wave}
\end{figure}

Electron dispersion in GaAs and graphene is isotropic under ambient condition and thus $a=1$. Now let us start by
considering a particularly simple case, namely a  small {\em uniform} anisotropy induced by either strain (in
either GaAs or graphene) or ripple (in graphene), that is possibly time-dependent:
\begin{eqnarray}
a=1+\xi(t).
\end{eqnarray}
This corresponds to space distortion induced by a long-wave length ``gravitational wave", in the gravity analogy.
Physically it can be induced by a lattice wave that is either of long-wave length, or with wave vector
perpendicular to the 2DEG plane so that the electrons see a uniform lattice distortion (see Fig. \ref{fig:Wave},
and more on this later).

We assume the frequency of $\xi(t)$ is low compared to Landau level spacing, thus no inter-Landau level
transition is induced. Then the main physical effect of this time-dependent geometry comes from its coupling to
intra-Landau level dynamics of the electrons.
This results in a time-dependent perturbation in the intra-Landau level Hamiltonian:
\begin{eqnarray}
\delta\tilde{V}(t)=\frac{\xi(t)}{4}\sum_{\bf q}(q^2_y - q^2_x)V_{\bf
q}e^{-\frac{1}{2}q^2\ell^2}\overline{\rho}_{\bf q}\overline{\rho}_{-{\bf q}}.
\label{eq:perturbation}
\end{eqnarray}
The 2DEG (assumed to be in its ground state) will absorb energy from the ``gravitational wave", with a rate
determined by the spectral function
\begin{eqnarray}
I(\omega)=\sum_n|\langle n|\hat{O}|0\rangle|^2\delta(\omega-\omega_n),
\label{eq:SpectralFunction}
\end{eqnarray}
where $|0\rangle$ is the ground state, $|n\rangle$ is an excited state with excitation energy $\hbar\omega_n$,
and
\begin{eqnarray}
\hat{O}=\sum_{\bf q}(q^2_y - q^2_x)V_{\bf q}e^{-\frac{1}{2}q^2\ell^2}\overline{\rho}_{\bf
q}\overline{\rho}_{-{\bf q}}
\label{eq:gravitycoupling}
\end{eqnarray}
describes the coupling between 2DEG in a FQH state to the lattice distortion/geometry.
It is interesting to note that the ${\bf q}$-dependence of the term being summed over above takes a d-wave form,
indicating $\hat{O}$ carries angular momentum $L=2$. It will thus couple to excitations with $L=2$, which is the
case for gravitons.

\begin{figure}
\includegraphics[width=0.45\textwidth]{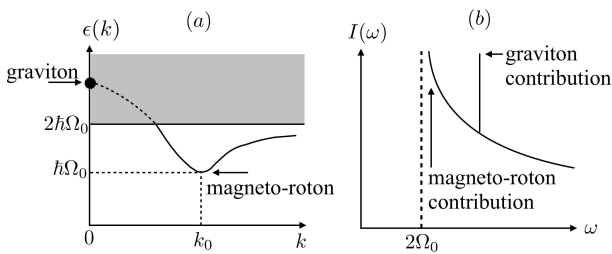}
\caption{(a) Illustration of excitation spectrum of Laughlin-type fractional quantum Hall states. Solid line
represents the magneto-roton mode, and shaded region represents two-roton continuum. The magneto-roton mode
continues into the two-roton continuum with decreasing wave vector $k$ (now represented as dashed line, all the
way to $k=0$, ending at the black dot that is the graviton mode which is the main focus of this paper. (b)
Spectral function of $\hat{O}$ defined in Eq. (\ref{eq:SpectralFunction}), revealing the presence of both the
graviton and magneto-roton modes.}
\label{fig:Spectrum}
\end{figure}

Excitation spectrum of Laughlin-type FQH states is illustrated schematically in Fig. \ref{fig:Spectrum}(a). The
lowest-energy elementary excitations are magneto-rotons (referred to as roton from now on) whose dispersion takes
the form
\begin{eqnarray}
\Omega(k)=\Omega_0 + A(k-k_0)^2,
\end{eqnarray}
where $\Omega_0$ is the (minimum) roton frequency (or roton gap), $k_0$ is the momentum of roton minimum, and $A$
is a constant. To a very good approximation\cite{GMP} a roton with momentum ${\bf k}$ is created by
$\overline{\rho}_{\bf k}$ for $k=|{\bf k}|\approx k_0$, and exhausts its spectral weight.
We thus find among other excitations, $\hat{O}$ creates a {\em pair} of rotons with total momentum zero, and the
roton pair contribution to the  spectral function $I(\omega)$ takes the form
\begin{eqnarray}
I_{roton}(\omega)&\propto&\sum_{|{\bf q}|\approx k_0}(q^2_y - q^2_x)^2 |V_{\bf q}|^2 e^{-q^2\ell^2}\langle
0|\overline{\rho}_{\bf q}\overline{\rho}_{-{\bf q}}|0\rangle \nonumber\\
&\times& \delta[\omega-2\Omega(q)],
\label{eq:RotonSpectralFunction}
\end{eqnarray}
It is easily seen that $I_{roton}(\omega)$ has a threshold frequency at $2\Omega_0$, and diverges for
$\omega\rightarrow 2\Omega_0 + 0^+$:
\begin{eqnarray}
I_{roton}(\omega)&\propto&\int{dq}\delta[\omega-2\Omega_0-2A(q-q_0)^2]\nonumber\\
&\propto&\frac{1}{\sqrt{\omega-2\Omega_0}}.
\end{eqnarray}
Thus the roton gap $\Omega_0$ is clearly visible in $I(\omega)$ [see Fig. \ref{fig:Spectrum}(b)]. This provides
an alternative method of measuring the magneto-roton gap, in addition to earlier attempts using optical
methods\cite{pinzuck,klitzing}.

What is more interesting, however, is the long-wavelength mode with $k\rightarrow 0$, which is the graviton mode
that is of primary interest of the present paper\cite{footnote}. It is known\cite{GMP,Haldane11,golkar} that
$\langle 0|\rho_{{\bf k}}\rho_{-{\bf k}}|0\rangle\propto (k\ell)^4$ as $k\rightarrow 0$, thus it is very
difficult to probe the collective mode in this regime using electromagnetic/optical probes that couple to
electron density. In particular the mode with ${\bf k}=0$ simply has no coupling to the ground state through the
density operator, as the cyclotron mode exhausts the spectral weight of the latter (Kohn's theorem). Another way
to understand this is the graviton has spin-2, and cannot be excited by spin-1 photons. On the other hand the
perturbation induced by acoustic/gravitational wave in Eq. (\ref{eq:perturbation}) is indeed an angular
momentum-two operator, and can excite the graviton mode (it is only natural that gravitons are excited by
gravitational waves!). We thus expect the graviton shows up as a {\em sharp} resonance in the spectral function
(\ref{eq:SpectralFunction}), allowing its energy to be measured spectroscopically [see Fig.
\ref{fig:Spectrum}(b)]. On-going numerical calculation of $I(\omega)$ defined in Eq. (\ref{eq:SpectralFunction}) indeed finds a pronounced peak corresponding to the graviton, which will be presented elsewhere\cite{RezayiYang}.

{\em Acoustic Wave as Gravitational Wave, and its other Effects} -- The experimental setup is schematically
illustrated in Fig. \ref{fig:Wave}. The acoustic waves propagate either inside the 3D bulk crystal or along its
surface, and interact with the mobile electrons that lives in a 2D layer through the lattice distortion or strain
they induce. There are several mechanisms for this (electron-phonon) interaction. As discussed above, the
strain-induced change of electron effective mass tensor corresponds to a geometric or gravitational interaction,
which is the main focus of the present work.
The specific geometric perturbation considered above corresponds to the setup of Fig. \ref{fig:Wave}(a), that
induces a {\em uniform} (but oscillating) strain in the 2D plane\cite{footnote1}.
There are, however, two other more familiar sources of coupling between strain and electrons which are likely to
be more important under generic situations\cite{benedict}: (i) The strain induces a deformation potential that
couples directly to the density of electrons. (ii) For non-centrosymmetric crystals, strain induces a electric
polarization and corresponding electric field due to the piezoelectric effect. We argue that the effects (i) and
(ii) may be eliminated by using the setup of Fig. \ref{fig:Wave}(a). In this case since the strain is {\em
uniform} in the 2D plane, the deformation potential is also uniform and thus has no effect. Similarly the
piezoelectric effect induces a {\em uniform} electric field, which couples to the {\em center-of-mass} of the
electron gas. Kohn's theorem guarantees that the dipole coupling of the electron gas to such uniform electric
field can only cause inter-Landau level transition, and no absorption will occur through such coupling as long as
$\omega < \omega_c$ (cyclotron frequency).

We now briefly consider the more generic case in which the lattice distortion is a function of both time and
space, with the latter dictated by a 2D wave vector ${\bf k}_0$. Such a distortion is induced by a 3D lattice
wave propagating with wave vector ${\bf K}_0$, whose 2D projection is ${\bf k}_0$ [Fig. \ref{fig:Wave}(b)], or by
a surface acoustic wave [Fig. \ref{fig:Wave}(c)]. In the gravity analogy we then have a (2D) ``gravitational
wave" with wave vector ${\bf k}_0$. However in these cases the other effects of (i) and (ii) mentioned in the
previous paragraph are generically present [although their effects are expected to vanish as $(k_0\ell)^4$], and
in the case of graphene non-uniform strain can also induce pseudo-gauge field. Thus additional effort is needed
to isolate the gravitational response. If this is possible, we can not only measure the graviton energy at zero
wave vector, but also its dispersion.

The acoustic wave absorption experiment proposed here has some similarity to earlier phonon absorption
experiments\cite{PhononAbsorpionExpt}. After all, an acoustic wave is made of coherent phonons or a phonon
version of laser. But there are a couple of fundamental difference here. (i) In earlier experiments phonons are
generated by a heat pulse, thus come with energies following a thermal distribution. This makes it impossible for
spectroscopy measurement. (ii) More importantly, since the thermal phonons come with un-controlled momenta and
polarizations, their effects are dominated by the deformation potential and piezo-electric polarization induced
by the strain\cite{benedict}. Here the wave vector and polarization of the acoustic wave are carefully controlled
so that the dominant effect of the strain is on the electron effective mass tensor or metric.

{\em Summary} -- In this paper we propose acoustic wave as an alternative probe of fractional quantum Hall
liquids, and demonstrate that it contains effects similar to those of gravitational wave. It allows for a direct
measurement of graviton energy, which is not possible using electromagnetic probes. While we focused on
Laughlin-type of states for their simplicity, the graviton as well as magneto-roton modes are expected to exist
in all fractional quantum Hall liquids, and can be probed using the methods described in this paper.

The authors (KY) has benefited from stimulating discussions with Lloyd Engel, Mansour Shayegan and Alexey
Souslov, and thanks Hsin-hua Lai and Mohammad Pouranvari for assistance.
This work was supported by DOE grant No. DE-SC0002140.


\begin{references}
\bibitem{gravitonnote} This is, of course, only an analogy to the quanta of space-time metric fluctuation in a quantum theory of (real) gravity. The reader is cautioned not to
    confuse between the two.
\end{references}
\end{document}